\documentclass{emulateapj}
\journalinfo{Accepted by ApJ on July 1, 2009}
\newcommand{\mbh}{M$_{\rm BH}$}
\newcommand{\redd}{r$_{\rm Edd}$}
\newcommand{\lya}{Ly$\alpha$}
\newcommand{\feii}{\ion{Fe}{2}}
\newcommand{\mgii}{\ion{Mg}{2}}
\newcommand{\civ}{\ion{C}{4}}
\newcommand{\nv}{\ion{N}{5}}
\newcommand{\ecs}{erg\,cm$^{-2}$\,s$^{-1}$}
\newcommand{\promille}{%
    \relax\ifmmode\promillezeichen
          \else\leavevmode\(\mathsurround=0pt\promillezeichen\)\fi}
  \newcommand{\promillezeichen}{%
    \kern-.05em%
    \raise.5ex\hbox{\the\scriptfont0 0}%
    \kern-.15em/\kern-.15em%
    \lower.25ex\hbox{\the\scriptfont0 00}}

\slugcomment{Accepted by ApJ on July 1, 2009}

\shorttitle{NIR spectroscopy of SDSS J0303-0019}
\shortauthors{Kurk et al.}

\begin{document}


\title{Near-infrared spectroscopy of SDSS J0303-0019:\\
       a low luminosity, high Eddington ratio quasar at $z \sim 6$
}


%
%
%


\author{J. D. Kurk\altaffilmark{1}, F. Walter\altaffilmark{1}, 
        X. Fan\altaffilmark{2,3}, L. Jiang\altaffilmark{2}, 
        S. Jester\altaffilmark{1}, H.-W. Rix\altaffilmark{1}, 
        D. A. Riechers\altaffilmark{4,5}}

\altaffiltext{1}{MPIA, K\"onigstuhl 17, 69117, Heidelberg, Germany}
\altaffiltext{2}{Steward Observatory, 
             933 N. Cherry Av, Tucson, AZ 85721}
\altaffiltext{3}{On leave at MPIA, Heidelberg, Germany}
\altaffiltext{4}{Caltech, 
                 1200 E.\ California Blvd,
                 Pasadena, CA 91125}
\altaffiltext{5}{Hubble Fellow}

\begin{abstract}
  We present sensitive near--infrared VLT ISAAC spectroscopic observations of
  the $z=6.08$ quasar SDSS J030331.40$-$001912.9.  This QSO is more than a
  magnitude fainter than other QSOs at $z\sim6$ for which NIR spectroscopy has
  been obtained to date and is therefore presumably more representative of the
  QSO population at the end of Cosmic Reionization.  Combining rest--frame UV
  continuum luminosity with the width measurements of the \ion{Mg}{2} and
  \ion{C}{4} lines, we derive a black hole mass of $2^{+1.0}_{-0.5}\times10^8$
  M$_\odot$, the lowest mass observed for $z\sim6$ QSOs to date, and derive an
  Eddington ratio of 1.6$^{+0.4}_{-0.6}$, amongst the highest value derived
  for QSOs at any redshift.  The Spitzer 24\,$\mu$m non--detection of
  this QSO does not leave space for a significant hot dust component in its
  optical/near--infrared SED, in common with one other faint QSO at $z=6$,
  but in contrast to more than twenty more $z=6$ QSOs and all known lower
  redshift QSOs with sufficiently deep multi-wavelength photometry.  We
  conclude that we have found evidence for differences in the intrinsic
  properties of at least one $z \sim 6$ QSO as compared to the lower--redshift
  population.
\end{abstract}

\keywords{Galaxies: high-redshift --- early Universe ---
             quasars: emission lines ---
             quasars: individual: SDSS J030331.40-001912.9
            }

\section{Introduction}

The radiation from luminous quasi stellar objects (QSOs) can currently be
studied up to redshift 6.5 and thus allows one to probe, over much of the
Universe's age, the properties of their central black holes \citep[BHs,
e.g.,][]{wil03}, their broad line region \citep[BLR, e.g.,][]{fre03}, and the
ionization state of intergalactic gas \citep[e.g.,][]{fan06c}.  The most
distant QSOs are of importance for both the study of galaxy evolution and
large--scale structure.  Among the most surprising discoveries in this field
are the detection of BH masses $>10^9$ M$_\odot$ \citep[e.g.,][]{bar03,wil03}
and the apparent constancy of the observed QSO BLR properties with redshift,
even up to $z = 6.4$ \citep[e.g.,][]{pen02}, corresponding to an age of
870\,Myr.  E.g., the \ion{Fe}{2}/\ion{Mg}{2} \citep[][hereafter
K07]{fre03,mai03,iwa04,kur07} and \ion{N}{5}/\ion{C}{4} \citep{jia07} emission
line ratios observed in the BLR of $z > 6$ QSOs are as high as that observed
in low-redshift objects.  Thus, near-solar metallicity in the centers of some
of the earliest systems and accretion of $>10^9$ solar masses in a central BH
can be attained in less than 1\,Gyr.  One caveat, however, has to be taken
into account: until now only the most luminous objects at $z \sim 6$ have been
studied.  These objects may experience faster evolution than less luminous
objects and therefore resemble low to intermediate redshift QSOs already at
$z\sim6$.  Secondly, the relationships established using low-luminosity AGN
(at low redshift) may not be valid for objects at the upper end of the
luminosity distribution.  As an example, the determination of the BH mass in
active galaxies was calibrated with reverberation mapping observations of AGN
that are one to two orders of magnitude fainter than the $z\sim6$ QSOs
observed until now \citep{kas00}.  It is therefore important to extend
$z\gtrsim6$ studies to lower--luminosity objects.  Here we present the first
observations of a QSO at $z>6$ with luminosity lower than L$_{\rm bol} =
10^{47}$ erg\,s$^{-1}$: the QSO \objectname{SDSS J030331.40-001912.9}
(hereafter J0303-0019).  This QSO was detected in the Sloan Digital Sky Survey
(SDSS) Deep Stripe with a magnitude of $z_{\rm AB} = 20.9$.  Its discovery
spectrum shows a very strong and narrow \lya\ emission line at $z=6.070$
\citep{jia08}.  As this line falls in the $z$-band, the actual continuum
luminosity is low, $m_{1450} = 21.3$, 1.5 magnitudes fainter than the mean of
the $z \sim 6$ QSO samples studied in K07 and \citet{jia07}.  In addition,
this is one of the two out of about twenty $z\sim6$ QSOs that are not detected
at 24\,$\mu$m with MIPS/Spitzer, implying a rest frame 3.5$\mu$m/4400\AA\
ratio far lower than for any other known QSO \citep[][L.\ Jiang et al. in
prep.]{jia06}.

\section{Observations and reduction}

\subsection{Near-infrared spectroscopy}

NIR spectroscopy of J0303-0019 in the $Y$ and $K$ bands was obtained in
service mode on December 24 and November 26 and 28, 2007, respectively, with
ISAAC \citep{moo98} at VLT UT1.  We used the short wavelength arm, equipped
with a 1024$\times$1024 Hawaii Rockwell array which has a pixel scale of
0.147\arcsec.  The 1\arcsec\ slit results in a spectral resolution of 550 and
450 for the $Y$ and $K$ bands, respectively.  Observations were carried out
in a standard ABBA offset pattern, with additional dithering to reduce the
influence of bad pixels.  Individual exposures were 148s, repeated 16 times
per observing block.  Nine such blocks were obtained in $K$ band, for a total
exposure time of $\sim$5.9\,hr, and two in $Y$ band, for a total time of
$\sim$1.3\,hr.
The data were reduced using ESO's ISAAC pipeline software, which does flat
fielding, background subtraction, rectification and combination of the
individual frames in one observing block.  We used sky lines to determine the
wavelength calibration, obtaining an absolute 1$\sigma$ error of 3 and 7\,\AA,
respectively, in $Y$ and $K$ band.  One-dimensional spectra were extracted
from the pipeline products in IRAF.  B and G stars with known NIR photometry
were used to correct for telluric absorption and initial flux calibration of
the one-dimensional spectra.  Finally, the one-dimensional spectra were
averaged to obtain the final spectra.  These were flux calibrated using the
broad band magnitudes obtained with the Calar Alto 3.5m telescope (see Sec.\
\ref{ssec:nir_img}).  The reduction is similar to that carried out by K07,
where more details are reported.

\subsection{Near-infrared broad band imaging\label{ssec:nir_img}}

For accurate flux calibration, the QSO SDSS J0303-0019 was imaged with the
prime-focus NIR camera Omega2000 \citep{kov04} at the 3.5m telescope at Calar
Alto, Spain.  This instrument has a 2k$\times$2k HAWAII-2 detector with a
pixel scale of 0.45\arcsec, resulting in a field of view of
15.4\arcmin$\times$15.4\arcmin.
J0303-0019 was observed on September 10 and 11, 2008, for 20, 20, 25, and
77 minutes in $Y, J, H$, and $K_s$ bands, respectively.
After dark subtraction and flat fielding with twilight exposures, the images
were sky subtracted, registered and averaged with XDIMSUM\footnote{XDIMSUM is
  a variant of the DIMSUM package developed by P.\ Eisenhardt, M.\ Dickinson,
  S.A.\ Stanford, and J.\ Ward.} in IRAF\footnote{IRAF is distributed by NOAO,
  which are operated by AURA, Inc., under cooperative agreement with NSF.}.
The seeing full width at half maximum (FWHM) on the combined images is
1.3\arcsec, 1.4\arcsec, 1.3\arcsec, and 1.0\arcsec, respectively.  The
combined images were astrometrized using the public service provided by {\tt
  astrometry.net} \citep{hog08} using a second order polynomial to correct for
distortions (which are small for Omega2000).  Subsequently, we used about 80
stars from the 2MASS All-Sky Catalog of Point Sources \citep{skr06} to obtain
the magnitude zero points.  For $Y$--band, we used the UKIDSS second data
release \citep{war07} catalog.  Only half of the observed field is covered
by UKIDSS, but this still provides 31 stars usable for the flux calibration.
Systematic errors in the flux calibration are estimated to be below 0.1\,mag.
The resulting fluxes \citep[using MAG\_BEST in SExtractor,][]{ber96} are
reported in Table~\ref{tbl:nir_mags}.

\begin{table}
\begin{center}
\caption{NIR Vega magnitudes of J0303-0019\label{tbl:nir_mags}}
\begin{tabular}{rrrr}
\tableline\tableline
\multicolumn{1}{c}{$Y$} & \multicolumn{1}{c}{$J$} & 
\multicolumn{1}{c}{$H$} & \multicolumn{1}{c}{$K_s$} \\
\tableline
20.60$\pm$0.14 & 20.44$\pm$0.08 &  19.78$\pm$0.08 & 18.95$\pm$0.09 \\
\tableline
\end{tabular}
\end{center}
\end{table}

\section{Results}

\subsection{Optical and NIR spectra}\label{ssec:spectra}

\begin{figure}
\plotone{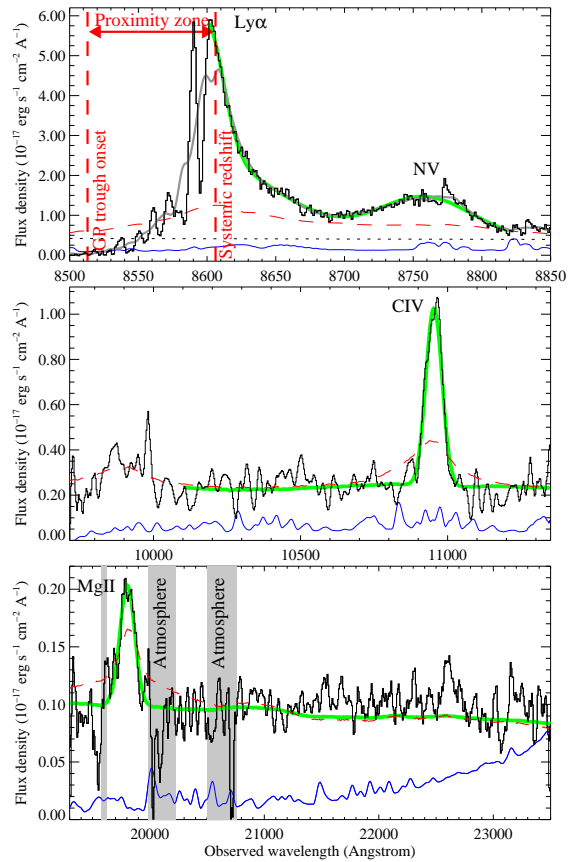}
\caption{{\it Top:} Optical spectrum of J0303-0019 (black histogram), the
  same spectrum smoothed over 20\,\AA\ (thick grey curve) and the noise per
  pixel (blue line).  A simultaneous fit (thick green curve) of the \nv\ line
  and the broad and narrow components of the \lya\ line is over plotted on top 
  of a power-law with slope of $\alpha_\lambda = -1.5$,
  normalized to a line free region of the spectrum (short-dashed line).  
  In addition, the wavelength of the onset of the GP
  trough and of \lya\ at the systemic redshift are indicated by
  vertical dashed lines.  The normalized SDSS QSO composite
  spectrum by \citet{vdb01} is over plotted in red for a visual comparison of
  the line widths.  {\it Middle and bottom:} NIR $Y$ and $K$ band spectra
  of J0303-0019, including the C\,IV and Mg\,II lines, respectively, smoothed
  over 5 pixels.
  The grey regions are seriously affected by sky lines and omitted from the 
  fit.  Lines and colors as in top panel.  The slope of the power-law continuum 
  here is $\alpha_\lambda = -1.0$.
  \label{fig:spectra}}
\end{figure}

\begin{table*}
\begin{center}
\caption{J0303-0019 Emission line measurements\label{tbl:line_meas}}
\begin{tabular}{rrrrrrrrrr}
\tableline\tableline
\multicolumn{1}{c}{Line} & $\alpha$  & \multicolumn{1}{c}{z} & 
\multicolumn{1}{c}{$\lambda_{\rm central}$} & 
\multicolumn{2}{c}{FWHM\tablenotemark{a}} &
\multicolumn{1}{c}{Flux\tablenotemark{b}} & \multicolumn{1}{c}{EW$_0$} &
\multicolumn{1}{c}{M$_{\rm BH}$} & 
\multicolumn{1}{c}{r$_{\rm Edd}$\tablenotemark{c}} \\ 
     &   &   & \multicolumn{1}{c}{\AA} & \multicolumn{1}{c}{\AA} & 
\multicolumn{1}{c}{km\,s$^{-1}$} &
\multicolumn{1}{c}{10$^{-17}$ cgs} & \multicolumn{1}{c}{\AA} &
\multicolumn{1}{c}{10$^{8}$ M$_\odot$} & \multicolumn{1}{c}{}  \\
\tableline
\lya\ (broad) &$-1.5$& 6.073$\pm$0.001& 8599$\pm$1& 125$\pm\:\;$5& 4350$\pm$165& 
                255$\pm\:\;$6  &   87$\pm$2\\
\lya\ (narrow)\tablenotemark{d}
              &      &          &           &  29$\pm\:\;$2& 1000$\pm\:\;$70&
                109$\pm$11     &   37$\pm$4\\
\nv           &      & 6.063$\pm$0.001& 8758$\pm$1&  81$\pm\:\;$2& 2760$\pm\:\;$80&
                 90$\pm\:\;$2  &   32$\pm$1\\
\tableline
\civ          &$-1.5$& 6.071$\pm$0.001&10953$\pm$1&  65$\pm\:\;$2& 1780$\pm\:\;$65&
                 54$\pm\:\;$2  &   37$\pm$1& 1.7$\pm$0.1  & 1.8$\pm$0.2 \\
\civ          &$-1.0$& 6.071$\pm$0.001&10953$\pm$1&  65$\pm\:\;$3& 1772$\pm\:\;$70&
                 54$\pm\:\;$2  &   36$\pm$1& 1.6$\pm$0.1  & 1.9$\pm$0.2 \\
\civ          &$ 0.0$& 6.071$\pm$0.001&10953$\pm$1&  64$\pm\:\;$3& 1760$\pm\:\;$70&
                 53$\pm\:\;$2  &   35$\pm$1& 1.5$\pm$0.1  & 2.0$\pm$0.2 \\
\tableline
\mgii         &$-1.5$& 6.078$\pm$0.001&19804$\pm$4&149$\pm$ 9& 2261$\pm$130&
                 16$\pm\:\;$1  &   25$\pm$1& 2.3$\pm$0.3  & 1.3$\pm$0.2\\
\mgii         &$-1.0$& 6.078$\pm$0.001&19803$\pm$4&154$\pm$ 8& 2340$\pm$125&
                 17$\pm\:\;$1  &   27$\pm$1& 2.5$\pm$0.3  & 1.2$\pm$0.1\\
\mgii         &$ 0.0$& 6.077$\pm$0.001&19801$\pm$3&167$\pm$ 8& 2525$\pm$120&
                 19$\pm\:\;$1  &   32$\pm$1& 3.0$\pm$0.3  & 1.0$\pm$0.1\\

\tableline
\end{tabular}
\tablenotetext{a}{Full width at half maximum in \AA\ (observed frame) and 
km\,s$^{-1}$ (rest frame)}
\tablenotetext{b}{Flux in 10$^{-17}$ \ecs}
\tablenotetext{c}{Eddington ratio: L$_{\rm bol}$ / L$_{\rm Edd}$ (CIV or MgII)} 
\tablenotetext{d}{Central wavelength fixed to that of the broad \lya\ line}
\end{center}
\end{table*}

In the top panel of Fig.~\ref{fig:spectra}, we show a close-up version of the
optical spectrum, first presented in \citet{jia08}, taken at the W.M.\ Keck
Observatory.  For comparison, the SDSS QSO composite spectrum published
by \citet[][dashed curve]{vdb01}, normalized to emission line free regions
of our observed spectra, is overlayed, showing that the emission lines of
J0303-0019 are much stronger and narrower than those present in the
composite spectrum.
We have made a simultaneous fit of the broad and narrow \lya\ and \ion{N}{5}
lines in the optical spectrum, after subtracting a standard \citep{vdb01}
power law continuum with a slope $F_\lambda \propto \lambda^{-1.5}$ as in
\citet{jia08}.  We fit only the spectrum redwards of the strong absorption
feature seen in the \lya\ line.  The best fit parameters are listed in
Table~\ref{tbl:line_meas}, and the best fit curves are over plotted in
Fig.~\ref{fig:spectra}.  The equivalent width (EW) and FWHM of \ion{N}{5}
measured by us are about 20\% higher than those reported by \citet{jia06},
who fixed the central wavelength of all fitted lines to the same redshift.

In the middle and bottom panels of Fig.~\ref{fig:spectra}, we show the VLT
ISAAC $Y$- and $K$-band spectra, containing the \civ\ and \mgii\ emission
lines, respectively.  There is also a tentative detection of
\ion{Si}{4}/\ion{O}{4}]1398 at 9900\,\AA\ in the $Y$-band, which we do not
try to fit.
The obtained NIR photometry of J0303-0019 (Table~\ref{tbl:nir_mags}) is
consistent with a power law SED with slope $\alpha_\lambda=-1.5$, as
measured in lower redshift QSOs \citep{vdb01}.  Imperfect flux calibration
of the NIR spectra, however, may affect the slope of the reduced $Y$ and $K$
band spectra.  Indeed, the slope observed in the $K$ band seems redder and
may be fit better by a slope of $\alpha_\lambda=-1.0$ or even
$\alpha_\lambda=0$.  As the NIR spectra are too short and the data not of
sufficient quality to constrain the slope, we carry out emission line fits
employing a range of slopes $\alpha_\lambda$ fixed to = 0, -1.0, and -1.5,
to cover the complete range of plausible slopes.  We take the slope of
$\alpha_\lambda=-1.0$ as the best compromise and use the parameters derived
from fits employing the two extreme values to construct conservative error
estimates on the fitted parameters.  The \civ\ and \mgii\ lines are fit by
single Gaussian functions on top of the co-addition of the power-law
continuum, a Balmer pseudo-continuum and a template of \feii\ lines
\citep[see also][]{kur07}.  The strength of the latter two components were
scaled with the power-law continuum, according to their proportions in the
composite QSO spectrum by \citet{vdb01}. As the strength of the \feii\
emission is not well constrained by the current spectra, but clearly lower
than four times the strength in the SDSS composite spectrum, we have also
carried out the fitting assuming a four times stronger and four times weaker
\feii\ contribution to establish proper error estimates.

The resulting best-fit parameters are displayed in
Table~\ref{tbl:line_meas}.  The \civ\ and \mgii\ lines have (rest
frame) FWHM$_0$ less than 2000 and 2600 km\,s$^{-1}$, and EW$_0$ of
36$\pm$1\,\AA\ and 27$_{-2}^{+5}$\,\AA, respectively.  The mean
FWHM$_0$ of these lines in the SDSS QSO sample of \citet[][hereafter
S08]{she08} is 5700 and 5850 km\,s$^{-1}$, respectively.  Only 2\%
and 1\% of the SDSS QSOs have \ion{C}{4} and \ion{Mg}{2} FWHM$_0$
smaller than 2000 and 2600 km\,s$^{-1}$, respectively.

\subsection{Black hole mass measurements}

The widths of the \ion{Mg}{2} and \ion{C}{4} lines, combined with an
independent estimate of their characteristic distance from the center, can be
used as a measure of the mass of the black hole (\mbh).  The relation between
line width, luminosity and \mbh\ has been calibrated empirically at low
redshift using reverberation mapping \citep[e.g.,][]{kas00}.  As in K07 and
\citet{jia07}, we use the relations for \ion{C}{4} and \ion{Mg}{2} presented
by \citet{ves06} and \citet{mcl04}, respectively, which are also those applied
to a sample of more than 60,000 SDSS QSOs by S08.  Due to the
  scatter of individual data points around this relation \mbh\ can be
  determined only within a factor of three.  For J0303-0019, we derive
\mbh\ with these methods and take into account the range of power-law
  slopes and \feii\ template strength described in Sec.~\ref{ssec:spectra}.
  We obtain \mbh\ of 1.6$_{-0.3}^{+0.1}$ and 2.5$_{-0.4}^{+0.5} \times$10$^8$
  M$_\odot$, respectively, from the \ion{C}{4} and \ion{Mg}{2} lines.  These
  are the lowest \mbh\ derived in a $z>6$ QSO to date, showing that, with
this sample of fainter $z \sim 6$ QSOs, we are indeed probing down the mass
function of BHs.  We note that the central wavelength of the \civ\ line is
blue shifted by 350 km\,s$^{-1}$ w.r.t.\ the \mgii\ line.  According to S08,
this modest shift ($\leq 1000$ km\,s$^{-1}$) implies that the \civ\ line
profile is dominated by gravitational effects and that \mbh\ can therefore be
reliably estimated from this line.

\begin{figure}
\epsscale{1.00}
\plotone{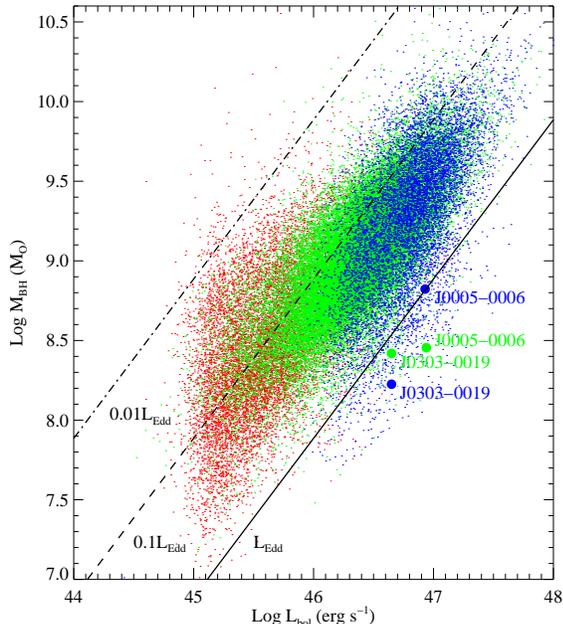}
\caption{The accretion luminosity of J0303-0019 and J0005-0006: the figure
  shows the M$_{\rm BH}$ -- L$_{\rm bol}$ diagram reproduced from S08,
  including $\sim$60,000 SDSS QSOs at $0.1 \le z \le 4.5$.  Colors show
  M$_{\rm BH}$ estimates using the following virial estimators: red for
  H$\beta$, green for Mg\,II, blue for C\,IV.  Three diagonal lines show 0.01,
  0.1, and 1 $L_{\rm Edd}$, as indicated.  Superposed are the data points for
  J0303-0019 (this work) and for J0005-0006 (K07).  Here, we have applied the
  cosmology specified in S08, resulting in M$_{\rm BH}$ (L$_{\rm bol}$) larger
  by 5\% (7\%), but \redd\ similar to those in our cosmology.
    \label{fig:mass_lum}}
\end{figure}

The \mbh\ derived from the line widths and corresponding Eddington
luminosities of $\log L_{\rm Edd}$ = 46.31$_{-0.09}^{+0.02}$, and
46.51$_{-0.07}^{+0.08}$ erg\,s$^{-1}$, respectively, are small
compared to the QSO's bolometric luminosity of $\log L_{\rm bol} =
46.59$$\pm$0.04 ergs\,s$^{-1}$, implying a high Eddingtion ratio
(\redd). From optical, IR, and radio data (L.\ Jiang et al, in prep.),
the bolometric luminosity of J0303-0019 is determined as in
\citet{jia06} to be $\log L_{\rm bol} = 46.59$ ergs\,s$^{-1}$.  It is
close to the values of $\log L_{\rm bol}$ = 46.62, and 46.65
erg\,s$^{-1}$ obtained using the monochromatic luminosities at
1350\,\AA\ and 3000\,\AA, respectively, and the bolometric correction
factors\footnote{These bolometric correction factors are
  multiplicative in flux, unlike the common definition in units of
  magnitude.} employed by S08 (BC$_{1350}$ = 3.81, BC$_{3000}$ =
5.15).  Reconciling these values, we obtain the estimated bolometric
luminosity stated above.  The Eddington ratio of most QSOs is between
0.01 and 1 and only 804 out of the 62,185 SDSS with \mbh\ determined
from their line--width have \redd $> 1$ (S08).  The Eddington ratios
for J0303-0019 derived from the \civ\ and \mgii\ lines are \redd\ =
1.9$_{-0.4}^{+0.2}$ and 1.2$_{-0.2}^{+0.3}$ (see
Table~\ref{tbl:line_meas} and Fig.~\ref{fig:mass_lum}).  Less than 2\%
and 0.2\% of the SDSS QSOs have such an \redd(\civ)$>$1.9 and
\redd(\mgii)$>$1.2.  Even if we assume that the other four (faint)
$z\sim6$ QSOs of the complete sample of QSOs discovered in the SDSS
Deep Stripe \citep{jia08} have \redd\ $<$1.2, the fraction of L$_{\rm
  bol} < 10^{47}$ erg\,s$^{-1}$ QSOs at $z\sim6$ with \redd\ $>$1.2 is
33\%, almost 40 times higher than at lower redshifts.

\subsection{Abundance indicators}

The chemical composition of gas clouds in the BLR can be explored by
comparing observed emission line ratios to models.  The heavy elements
in the BLR are products of the star formation history in the host
galaxy and therefore indicators of past star formation in at least the
very center of the host.  \citet{nag06} present several emission line
ratios observed in SDSS QSOs at $2.0 \ge z \ge 4.5$ and models
explaining the ratios.  According to \citeauthor{nag06}, the strength
of the \ion{N}{5}/\ion{C}{4} ratio is strongly correlated with the
luminosity of the QSO, which can be interpreted in terms of higher gas
metallicity in more luminous QSOs.  Although J0303-0019 is one of the
faintest QSOs at $z\sim6$, it has a $M_B$ of -26.2, which makes it
moderately luminous in comparison with the SDSS QSOs studied by
\citeauthor{nag06}.  We measure a ratio of \ion{N}{5}/\ion{C}{4} =
1.61$\pm$0.07, the highest ratio among the known $z \sim 6$ QSOs, and
also higher than those measured in the composite spectra of quasars in
the luminosity range $-25.5 > M_{\rm B} > -26.5$ by \citet{nag06}.
This suggests that the BLR has a high metal abundance, possibly as
high as 10Z$_\odot$ \citep[see Fig.~29 in][]{nag06}, but the authors
warn that estimates of the BLR metallicity using only emission line
ratios involving \ion{N}{5} might be quite uncertain \citep[see
also][]{ham02}.

\subsection{Str\"omgren sphere}

The intense UV radiation of the QSO keeps a region around the QSO
ionized, the Str\"omgren sphere, which will actually only be spherical
if the IGM surrounding the QSO is homogeneous.  Outside this region,
UV photons are absorbed by neutral hydrogen and this causes complete
absorption blueward of the wavelength of redshifted \lya, the
so-called GP trough \citep{gun65}.  For reasons discussed in
\citet{fan06c}, the Str\"omgren sphere radius is difficult to measure
and we rather use the size of the proximity zone (R$_{\rm P}$),
defined as the region in which the transmitted flux is $>$10\% of the
(extrapolated) continuum emission when smoothed to a resolution of
20\,\AA\ (as shown by the grey line in Fig.~\ref{fig:spectra}).  We
measure the blue edge of this region to be at 8513$\pm$1\,\AA\ (dashed
line in Fig.~\ref{fig:spectra}).  Combined with the systemic redshift
of $z=6.078$ derived from the \mgii\ line (redmost dashed line in
Fig.~\ref{fig:spectra}), we compute (K07) the co--moving radius of the
Str\"omgren sphere to be R$_{\rm S}$ = 4.5$\pm$0.3\,Mpc.  This is a
typical value for luminous $z\sim6$ QSOs \citep[see, e.g.,
K07,][]{wal03,fan06d}.  The average size of QSO Str\"omgren spheres,
however, decreases by a factor $\sim$2.5 over $5.7 < z < 6.4$,
consistent with an increase of the neutral fraction with redshift
\citep{fan06c}.

\section{Discussion and conclusion}

SDSS J0303-0019 has the lowest BH mass known at $z\sim6$ and is a
truly peculiar QSO: it has one of the highest Eddington ratios
observed for a QSO, it has the highest \ion{N}{5}/\ion{C}{4} ratio
observed in $z\sim6$ QSOs, which is also significantly higher than
that observed in lower redshift composite spectra of QSOs, and it is
one of the two $z\sim6$ QSOs that is not detected in deep 24\,$\mu$m
observations \citep[][L.\ Jiang et al.\ in prep.]{jia06}.  The other
QSO which is not detected at 24\,$\mu$m is \objectname{SDSS
  J000552.34-000655.8} (hereafter J0005-0006), which also has a high
Eddington ratio: 2.4 as derived from the \ion{Mg}{2} based \mbh\ and
1.0 as derived from the \ion{C}{4} based \mbh (K07).  Detections at
250\,GHz and 1.4\,GHz have also been attempted but neither of the
objects were detected down to rms values of 0.51, 0.48\,mJy and 62,
130\,$\mu$Jy, respectively \citep{wan08}.  In contrast, these QSOs
were detected at 3.6, 4.5 and 5.8\,$\mu$m with IRAC/Spitzer
\citep[][L.\ Jiang et al.\ in prep.]{jia06}, implying a rest frame
3.5$\mu$m/4400\AA\ ratio far lower than for any object in the $z\sim6$
QSO sample, and lower than for any of the PG QSOs observed by
\citet{neu87} and \citet{haa00,haa03} for which the sensitivity at
3.5mu rest-frame is not an issue.  \citet{jia06} show that the
optical/IR SEDs of the observed $z\sim6$ QSOs can be fit by two
components: a power-law representing an accretion disk and a
black-body representing hot dust.  The respective (1\,$\sigma$) upper
limits of 21\,$\mu$Jy and 9\,$\mu$Jy at 24\,$\mu$m for J0303-0019 and
J0005-0006, significantly lower than the expected flux of
$\sim$160\,$\mu$Jy based on SEDs of other $z\sim6$ QSOs, suggests the
absence of hot dust.  The implied lack of dust could be explained if
the gas is metal--poor, as in a galaxy host which has not experienced
major star formation yet.  However, this interpretation would
contradict the potentially high metallicity based on the measured
\ion{N}{5}/\ion{C}{4} ratio. Alternatively, the non-detection of the
hot dust component could be due to different dust properties, or due
to a surrounding cool dust torus hiding the hot component
\citep{jia06}.

We conclude that, with the observation of QSOs with bolometric luminosity
$\lesssim 10^{47}$ erg\,s$^{-1}$, we start to probe a population of QSOs that
shows different observed properties as compared to higher luminosity $z\sim6$
QSOs and lower--redshift QSOs of all luminosities.  We can only speculate that
the hosts of the lower--luminosity QSOs at $z\sim6$ may be younger than
higher--luminosity QSOs at the same redshift because the host galaxies of the
latter are more massive and have therefore formed earlier.  It is, however,
unclear at this stage what physical mechanisms are causing the observed
differences.

\acknowledgments

We thank the anonymous referee for comments that helped to improve the
paper.  Based on observations carried out at the European Southern
Observatory, Chile (Program No.\ 080.A-0794) and at the Calar Alto
Observatory, Spain.  JK acknowledges financial support from
\emph{Deutsche Forschungsgemeinschaft} (DFG) grant SFB 439.  XF and LJ
acknowledge supports from NSF grants AST 03-07384 and AST 08-06861 and
a Packard Fellowship for Science and Engineering. XF acknowledges
additional support from the Max Planck Society and a Guggenheim
Fellowship.  DR acknowledges support from NASA through Hubble
Fellowship grant HST-HF-01212.01A awarded by the STScI, operated by
AURA, under contract NAS 5-26555.  This publication makes use of data
products from the Two Micron All Sky Survey, which is a joint project
of the University of Massachusetts and the Infrared Processing and
Analysis Center/California Institute of Technology, funded by the
National Aeronautics and Space Administration and the National Science
Foundation.


{\it Facilities:} \facility{VLT:Antu}, \facility{CAO:3.5m ()}

\bibliographystyle{apj}



\end{document}